\documentclass[doublespacing]{elsart}
\newcommand{\scs}{\scriptscriptstyle}
\newcommand{\tsi}{\tilde{\sigma}}

\usepackage{amssymb}

\begin{document}

\begin{frontmatter}



\title{Mass transfer in field of fast-moving deformation disturbance}


\author{G.L. Buchbinder}
\ead {buchb@univer.omsk.su}
\address{Department of Physics, Omsk
State University, Peace Avenue 55-a 644077, Omsk, Russia}

\begin{abstract}
The mass transfer of interstitial impurities in a crystalline
lattice under the influence  of the fast-moving deformation
disturbance of the type of a shock wave is considered. The
velocity of the movement of the disturbance is supposed to be
compared with the characteristic velocity of the relaxation of the
diffusion flux to its local equilibrium value determined by the
Fick's law. The similar situation occurs in a number of
experiments on the exposure of a solid to dynamical external loads
giving rise to such fast hydrodynamical processes in a sample that
the local equilibrium assumption, normally assumed for the
macroscopic description  of transport processes, is no longer
valid. Considering the diffusion flux among the set of independent
variables we have derived a set of coupled hydrodynamic equations
describing nonequilibrium behavior of a solid in the absence of
local equilibrium in the system. Within the scope of the proposed
model it has been shown that in comparison with the local
equilibrium system an enhanced mass transfer occurs under local
nonequilibrium conditions.

\end{abstract}

\end{frontmatter}

\twocolumn[\centering \bf  NOMENCLATURE \vspace{2cm}]
$c$  mass concentration of interstitialcs\\
$E$ energy density\\
${\bf J}$ diffusion flux density of interstitials\\
$p$ pressure\\
${\bf Q}$ energy flux density \\
${\bf q}$ heat  flux density\\
$S$ entropy density of the medium\\
$T$ temperature\\
$t$  time\\
${\bf u}$ displacement vector\\
$ u_{ij}$ infinitesimal strain tensor\\
$V$  velocity of the deformation disturbance\\
$V_D$ speed of the propagation of concentration perturbations\\
\vspace{8cm}

${\bf v}$ mass velocity\\
$x_i$ coordinate\\

{\large Greek symbols} \\
$\delta$ width of the transition region\\
$\mu$ chemical potential of the medium\\
$\nu$  chemical potential of the interstitials per unit of
volume\\
$\xi$ dimensionless coordinate\\
$\Pi_{ik}$ the tensor momentum flux density\\
$\pi_{ij}$  viscous pressure tensor \\
$\rho$  total mass density \\
$\sigma_{ij}$ elastic stress tensor\\
$\tau$ relaxation time of the diffusion flux\\
$\varphi = V/V_D$ constant value\\
\onecolumn
\section{Introduction}
\label{sec:1} In recent experiments \cite{K94,BV99} on the
irradiation of the thin-film metal samples by high power ion beams
(HPIB) thin films of aluminium and silver on the niobium and
copper substrates were irradiated by carbon ions. The considerable
penetration depth of the film atoms into the substrate, exceeding
the length of the projective path of carbon atoms, has been
revealed in the irradiated samples. The estimation of the
effective diffusion coefficients has given the anomalously high
magnitudes \cite{K94,BV99,BK99}  $D \sim 10^{-2} - 10^{-4}$
sm$^{2}$s$^{-1}$. The effect of the intensive mass transfer by
HPIB irradiation has been registered earlier in work \cite{P90}. A
similar effect has also been observed at electron and laser
irradiation of solids \cite{BG83,BC86,NT86} and at shock loading
of the samples \cite{MP84,NV85}. The complete explanation of this
phenomenon within the scope of the known diffusion mechanisms has
not yet been found.\\
One of possible causes responsible  for the accelerated mass
transport is the shock wave, generated by irradiation, which is
able to induce the enhanced migration of atoms on large depths
\cite{BV99}. In investigating this mechanism of diffusion within
the scope of the hydrodynamic description usually one starts from
the known diffusion equation in which, according to Fick's law,
the diffusion flux is defined by gradients of the concentration of
a diffusing substance, a pressure and a temperature. Such approach
is well justified if the disturbances occurring in a system change
slowly enough in comparison with the characteristic rate at which
the given system approaches the local equilibrium. Meanwhile the
interaction of particle beams with a solid induces in the latter
rather fast hydrodynamic processes with the characteristic times
of order or less than $10^{-8}$s and the shock velocity can reach
several kilometers per secund. Under these conditions one should
expect a significant deviation of the system state from local
equilibrium and the violation of Fick's law and take into account
an influence of the relaxation of the system on the process of
mass transport. \\
In this paper we want to study within a simple model basic
features the impurity transfer in the relaxating solid medium.
More specifically the purpose of the work is to research the
interstitial diffusion in the field of the fast moving deformation
disturbance of the type of the shock wave. The velocity of the
movement of this disturbance is supposed to be compared with the
velocity  of the propagation of concentration disturbances in the
medium. The latter condition just defines the condition of absence
of local equilibrium in the system \cite{S97}. The task is also
interest in connection with shock-compression experiments with
metals containing considerable atomic concentrations of
interstitial
hydrogen \cite{TF94,PS99}.\\
The description of the fast hydrodynamical processes requires an
extended number of independent variables, defining the
nonequilibrium  state of a medium, in comparison with conventional
hydrodynamics. During the course  of short enough the
characteristic time intervals, in particular, the diffusion flux
may not have time to relax to its local-equilibrium value, defined
by Fick's law and it should be considered as independent variable
obeying some relaxation equation. The dissipative fluxes
(diffusion flux, viscous pressure tensor, heat flux) included in
the local conservative laws of mass, momentum and energy are
frequently exploited as additional independent variables
\cite{JL96,S97}.\\
In section\ \ref{sec:2} we determine the total set of the
independent variables suitable for the description of the fast
hydrodynamical processes in the crystalline solid with
interstitials. For these variables the derivation of the set of
the coupled hydrodynamical equations, including a relaxation
equation for the diffusion flux is given. Section\ \ref{sec:3} is
devoted to the investigation of the mass transport in the field of
fast-moving deformation disturbance of the crystal in the absence
of local equilibrium in the system. Resume and concluding remarks
are given in section\ \ref{sec:4}.

\section{Derivation of hydrodynamic equations}
\label{sec:2}

In this section we want to derive the set of equations relevant
for the description of fast hydrodynamical processes  in the
crystalline solid with interstitial impurities. The complete  set
of the hydrodynamical equations has to contain the local
conservation laws of mass , momentum and the equation for the
entropy production
\begin{eqnarray}
&&\frac{\partial\rho}{\partial t} + \mbox{div}\,\rho {\bf v}  =  0\, ,\label{1}\\
&&\frac{\partial (\rho v_i)}{\partial t} + \frac{\partial\Pi _{ik}}{\partial x_k}  =  0\, ,\label{2}\\
&&\frac{\partial S}{\partial t} + \mbox{div}\, (S{\bf v} +
\frac{{\bf q}}{T} - \frac{\nu}{\rho T}\,{\bf J}) = \frac{R}{T}\, .
\label{3}
\end{eqnarray}
Here $\rho$ is the total mass density of the medium (lattice plus
interstitials), ${\bf v}$ is the mass velocity, $S$ is the entropy
density of the medium, $\Pi_{ik}$ is the tensor momentum flux
density, ${\bf q}$ is the heat  flux density, ${\bf J}$ is the
diffusion flux density of interstitials, $\nu$ and $T$ are the
chemical potential of the interstitials per unit of volume and the
absolute temperature, respectively, and $R\, (R > 0)$ is the
dissipative function of the medium; summation over repeated Latin
indices is implied.\\
Further we suppose that a concentration of vacancies is negligible
or their mobility is significantly less than  the mobility of
interstitials  so that the mass transport occurs mainly by the
interstitial motion. Introducting the mass concentration of the
interstitial particles $c$ we write down the continuity equation
for the interstitial density $c\rho$ as follows
\begin{eqnarray}
\frac{\partial(c\rho )}{\partial t} + \mbox{div}\,( c\rho {\bf v}
+ {\bf J}) =  0\, .\label{4}
\end{eqnarray}

It should be noted that the diffusion flux ${\bf J}$, appearing in
equations (\ref{3}) and (\ref{4}), is determined in a rest frame
(moving with velocity ${\bf v}$ relative to the laboratory frame).
Meanwhile for a crystal it is convenient to determine the
diffusion flux ${\bf J}_{\scs L}$ relative to a lattice. It is
easy to see that both fluxes are related by ${\bf J} = (1 - c){\bf
J}_{\scs L}$. As the mass concentration $c \sim m_p/m_{\scs L}$,
where $m_p$ and $m_{\scs L}$ are masses of an impurity particle
and a host atom, then for light impurities (type of hydrogen),
occurring into a  matrix of heavy atoms, $c << 1$ even for the
finite atomic concentration. Because of this ${\bf J} \simeq {\bf
J}_{\scs L}$.\\
The variables $\rho, {\bf v}, c, S$, (or $T$) define conventional
set of the independent hydrodynamical variables for a "fluid-like"
medium. A solid with point defects has the extra independent
hydrodynamical variable. This is a displacement vector ${\bf u}$
defining displacement of lattice sites \cite{Fl76}. In addition,
according to above said, in the case of fast processes the
dissipative fluxes - diffusion flux ${\bf J}$, heat flux ${\bf q}$
and viscous pressure tensor $\pi_{ik}$ must be included among the
set of independent variables. For simplicity we shall assume that
the relaxation times of fluxes ${\bf q}$ and $\pi_{ik}$ are
considerably less than the relaxation time of diffusion flux and
during characteristic time of the process under consideration  the
fluxes ${\bf q}$ and $\pi_{ik}$ have time to relax to their local
equilibrium values \cite{S97}. In this case diffusion flux ${\bf
J}$ is the only new independent variable.\\
Thus we assume that quantities $\rho, {\bf v}, c, S, {\bf u}, {\bf
J}$ completely define the nonequilibrium state of the system. For
such set variables the set of the hydrodynamical  equations
(\ref{1})-(\ref{4}) has to be supplemented by the equations for
the displacement vector ${\bf u}$ and the diffusion flux ${\bf
J}$. To derive these equations and define the still unknown
$\Pi_{ik}, {\bf q}$ and $R$, appearing in equations
(\ref{1})-(\ref{4}), we use the method originally applied in the
theory  of the superfluids for the derivation  of the equations of
the two-fluid hydrodynamics \cite{LL87}. The main idea, in the
given case, is as follows. We shall find these quantities so that
the conservation energy law
\begin{equation}
\frac{\partial E}{\partial t} + \mbox{div}\,{\bf Q}  =  0\, ,
\label{5}
\end{equation}
would follow from equations (\ref{1})-(\ref{4}) and the equations
for $\dot{{\bf u}},\, \dot{{\bf J}}$ (an upper dot denotes the
time derivative). In the equation (\ref{5}) $E$ and $\bf Q$ are
the energy density and the energy flux density of the medium,
respectively.\\
The energy $E$ is related by the Gallilean transformation to its
value  $E_0$ in the frame, where a given element of the volume of
the medium rests, by the relationship
\begin{equation}
E = E_0 + \frac{\rho{v}^2}{2}\, . \label{6}
\end{equation}
Let us now write down the differential of $E_0$, considered as a
function  of $S, \rho , c, \bf J$ and the infinitesimal strain
tensor $u_{ij}$ in the form
\begin{equation}
dE_0 = TdS + \mu d\rho + \nu d c +
\tilde{\sigma}_{ij}d\tilde{u}_{ij} + {\bf w}\cdot d\,{\bf J} \, ,
\label{7}
\end{equation}
here $\mu$ is the chemical potential of the medium, ${\bf w} =
{\partial E_0}/{\partial {\bf J}}$ is the conjugate variables with
$\bf J$, and $\sigma_{ij}$ is the symmetric, elastic stress
tensor; here and further the tilde is used to denote the traceless
part of a tensor. In the equation (\ref{7}) it has been taken into
account that the variation $du_{ii}$ of sum of diagonal components
of the strain tensor $u_{ij}$ is determined by the variation of
the density  $d\rho$.\\
Differentiating the equation (\ref{6}) with respect to time and
using equation (\ref{7}), one obtains
\begin{equation}
\frac{\partial E}{\partial t} = {\frac{\partial}{\partial t
}}\Bigl(\frac{\rho v^2}{2}\Bigr) + T\frac{\partial S}{\partial t}
+ \mu\frac{\partial\rho}{\partial t} + \nu\frac{\partial c
}{\partial t} + \tsi_{ij}\frac{\partial{\dot u}_{i}}{\partial x_j}
+ w_i\,{\dot J}_i\, . \label{8}
\end{equation}
Employing equations (\ref{1})-(\ref{4}) we can write down the time
derivatives as
\begin{eqnarray}
&&\frac{\partial }{\partial t}\Bigl(\frac{\rho v^2}{2}\Bigr)= -
\frac{v^2}{2}\frac{\partial\rho}{\partial t} + v_i\frac{\partial
(\rho v_i) }{\partial t}  \nonumber \\
&&\phantom{\frac{\partial }{\partial t}\Bigl(\frac{\rho
v^2}{2}\Bigr)} =- \mbox{div}\Bigl(\frac{\rho v^2}{2}{\bf v}\Bigr )
-
v_i\frac{\partial }{\partial x_k}(\Pi_{ik} - \rho v_i v_k)\, , \label{9} \\
&&T{\dot S} =-T\mbox{div}\Bigl (S{\bf v} + \frac{{\bf q}}{T} -
\frac{\nu}{\rho T}\,{\bf J} \Bigr ) + R \nonumber \\
&&\phantom{T{\dot S}}= - \mbox{div}\Bigl (TS{\bf v} + {\bf q} -
\frac{\nu}{\rho}\, {\bf J } \Bigr ) + {\bf v}\cdot\nabla (TS) -
T{\bf v}\cdot\nabla S  \nonumber\\
&&\phantom{T{\dot S}\hspace{0.7cm}=}- {\bf J}\cdot\nabla\Bigl
(\frac{\nu}{\rho}\Bigr ) + R + {\bf q}\cdot\frac{\nabla T}{T} +
T{\bf J}\cdot\nabla\Bigl (\frac{\nu}{\rho T}\Bigr )\, ,\label{10}\\
&&\nu\dot{c} = -\nu{\bf v}\cdot\nabla c - \mbox{div}\Bigl (
\frac{\nu}{\rho}\,{\bf J}\Bigr ) +  {\bf
J}\cdot\nabla\Bigl (\frac{\nu}{\rho}\Bigr )\, ,\label{11}\\
&&\mu\dot{\rho} = - \mbox{div}(\mu\rho {\bf v}) + {\bf
v}\cdot\nabla
(\mu\rho) - \mu {\bf v}\cdot\nabla\rho \, , \label{12}\\
&&\tsi_{ij}\frac{\partial \dot{u}_i}{\partial x_j} =
\frac{\partial}{\partial x_j}(\tsi_{ij}\dot{u}_i) -
\dot{u}_i\frac{\partial\tsi_{ij}}{\partial x_j}\, . \label{13}
\end{eqnarray}
Inserting the equations (\ref{9})-(\ref{13}) into
equation(\ref{8}) and taking into account that
\[\nabla E_0 = T\nabla S + \mu\nabla\rho + \nu\nabla c  +
\tsi_{ij}\nabla\Bigl (\frac{\partial u_i}{\partial x_j} \Bigr ) +
w_i\nabla J_i\, , \] after long, though simple, transformations,
one has
\begin{eqnarray}
\dot{E} = - \mbox{div}\Bigl\{\frac{\rho v^2}{2}{\bf v} + TS{\bf v}
+ \mu\rho{\bf v} + {\bf q} - \dot{\bf u}\cdot{\bf \tsi}\Bigr\} +
v_k\tsi_{ij}\frac{\partial}{\partial x_j}\Bigl(\frac{\partial
u_i}{\partial x_k}\Bigr)&&\nonumber \\
 - v_i\frac{\partial}{\partial x_k}\bigl(\Pi_{ik} - \rho v_iv_k
+ \tsi_{ik} \bigr) + {\bf v}\cdot\nabla (- E_0 + TS + \mu\rho)&&\nonumber\\
  + R + {\bf q}\cdot\frac{\nabla T}{T} + T{\bf
J}\cdot\nabla\Bigl(\frac{\nu}{\rho T}\Bigr) + (v_i -
\dot{u}_i)\frac{\partial \tsi_{ij}}{\partial x_k}&& \nonumber\\
 + w_i (\dot{J}_i + {\bf v}\cdot\nabla J_i)&&\label{14}
\end{eqnarray}
Finally equation (\ref{14}) can be rewritten as
\begin{eqnarray}
\dot{E} + && \mbox{div}\{\frac{\rho v^2}{2}{\bf v} + TS{\bf v} +
\mu\rho{\bf v} + {\bf q} - \dot{\bf u}\cdot{\bf \tsi} + {\bf v
}\cdot{\bf \pi} \Bigr\} \nonumber\\
 = & & R + {\bf q}\frac{\nabla T}{T} +  \pi_{ik}\frac{\partial
v_i}{\partial x_k} + T{\bf J}\cdot\nabla\Bigl(\frac{\nu}{\rho
T}\Bigr)
+ (v_i - \frac{d\,u_i}{d\,t})\frac{\partial \tsi_{ik}}{\partial
x_k} + w_i\frac{d J_i}{d\,t}\, , \label{15}
\end{eqnarray}
where the viscous pressure tensor $\pi_{ik}$ is defined from the
equality
\begin{equation}
\Pi_{ik} = \rho v_iv_k + p\delta_{ik} - \tsi_{ik} + \pi_{ik}
\label{16}
\end{equation}
and the notations have been introduced
\begin{equation}
p = - E_0 + TS + \mu\rho \label{17},
\end{equation}
and for the material derivative $d/d\,t = {\partial}/{\partial t}
+ {\bf v}\cdot\nabla $ . \\
In the derivation equation (\ref{16}) we
have neglected the term of the second order of infinitesimal in
strains $\tsi_{jk}\partial
u_i/\partial x_j$ in comparison with the linear term $\tsi_{ik}$.\\
Equation (\ref{16}) defines the momentum flux density, where  $p$
can be interpreted as "pressure"  in the medium. Comparing further
equation (\ref{15}) with the energy conservation law (\ref{5}) we
define the energy flux density  ${\bf Q}$ and the dissipative
function $R$ as
\begin{equation}
{\bf Q} =  \Bigl(\frac{\rho v^2}{2} + E_0 + p \Bigr){\bf v} + {\bf
q} - \dot{\bf u}\cdot{\bf \tsi} + {\bf v }\cdot{\bf \pi}\label{18}
\end{equation}
and
\begin{eqnarray}
R &=& - {\bf q}\cdot\frac{\nabla T}{T} -  \pi_{ik}\frac{\partial
v_i}{\partial x_k} - T{\bf J}\cdot\nabla\Bigl(\frac{\nu}{\rho
T}\Bigr) \nonumber\\
& &\hspace{4cm}- (v_i - \frac{d\,u_i}{d\,t})\,\frac{\partial
\tsi_{ik}}{\partial x_k} - w_i\frac{d J_i}{d\,t}\, , \label{19}
\end{eqnarray}
Since the energy $E_0$ is even under time reversal, the dependence
between $E_0$ and $J_i$ in the simplest approximation has the form
\[E_0 = \frac{1}{2}\,W_{ij}J_iJ_j + \ldots\, ,\hspace{1cm}W_{ij} = W_{ji}\, ,\]
where $W_{ij}$ is assumed to be constants and dots denote the
terms of the highest order of infinitesimal in fluxes. In this
case
\begin{equation}
w_i = \frac{\partial E_0}{\partial J_i} = W_{ik}J_k \label{20}
\end{equation}
and the dissipative function $R$ can be rewritten as
\begin{equation}
R = - {\bf q}\cdot\frac{\nabla T}{T} -  \pi_{ik}\frac{\partial
v_i}{\partial x_k} - (v_i - \frac{d\,u_i}{d\,t})\,\frac{\partial
\tsi_{ik}}{\partial x_k} - J_iX_i\, , \label{21}
\end{equation}
where
\[X_i = T\frac{\partial}{\partial x_i}\Bigl(\frac{\nu}{\rho
T }\Bigr) + W_{ij}\frac{dJ_j}{d\,t}\, .\] In the framework of the
linear theory the positive definiteness of $R$ leads to the linear
relationships relating the dissipative fluxes  to the thermodynamic
forces. Taking into account Onsager's reciprocity relations for the
transport coefficients and time-reversal property of dissipative
effects \cite{M72}, we can write these relationships in the form
\begin{eqnarray}
&&\pi_{ik} = - \eta_{iklm}\frac{\partial v_l}{\partial x_m}\, ,
\label{22}\\
&& q_i = - \frac{\kappa_{ik}'}{T^2}\frac{\partial T}{\partial x_k} -
\alpha_{ik}'X_k - \beta '_{ik}\frac{\partial
\tsi_{kj}}{\partial x_j}\, , \label{23}\\
&& J_i = - \frac{\alpha_{ik}'}{T^2}\frac{\partial T}{\partial x_k}
- \gamma_{ik}'X_k - \zeta '_{ik}\frac{\partial
\tsi_{kj}}{\partial x_j}\, , \label{24}\\
&& v_i - \frac{d u_i}{d\, t} = -
\frac{\beta_{ik}'}{T^2}\frac{\partial T}{\partial x_k} -
\zeta_{ik}'X_k - \chi '_{ik}\frac{\partial \tsi_{kj}}{\partial
x_j}\, , \label{25}
\end{eqnarray}
where the forth-rank tensor $\bf\eta$ is related to the effect of
the viscosity and $\alpha'_{ik}, \beta'_{ik}, \ldots$ are tensor
transport coefficients.\\
Let us eliminate the thermodynamic force $X_k$ from equations
(\ref{23}) and (\ref{25}). Redenoting the transport coefficients
and introducing  the explicit definition of $X_k$ into equation
(\ref{24}) one rewrites the set (\ref{23})-(\ref{25}) for the case
of the cubic lattice as
\begin{eqnarray}
&& {\bf q} = \alpha\, {\bf J} - \kappa\frac{\nabla T}{T^2} -
\beta\nabla\cdot{\bf \tsi }\, , \label{26}\\
&&\frac{d{\bf u}}{d\, t} = {\bf v} - \zeta\, {\bf J} +
\beta\frac{\nabla T}{T^2} + \kappa\nabla\cdot{\bf \tsi}\,
,\label{27}\\
&&\tau\frac{d{\bf J}}{d\,t} + {\bf J} = -\, \gamma
T\nabla\Bigl(\frac{\nu}{\rho T}\Bigr) - \omega\frac{\nabla T
}{T^2} - \lambda\nabla\cdot{\bf \tsi}\, , \label{28}
\end{eqnarray}
where $\alpha, \beta, \gamma\ldots$ are the scalar transport
coefficients. $\kappa$ is connected with the heat conductivity ,
$\tau$ is the relaxation time of the diffusion flux. The
coefficients $\beta, \kappa, \lambda$ determine the relation
between the dissipative effects and shear stresses. The remaining
coefficients are due to the processes of diffusion, thermal
diffusion and barodiffusion  (owing to the pressure dependence of
the interstitial chemical potential $\nu$ in equation
(\ref{28})).\\
Thus the complete set of the hydrodynamic equations of the system
under consideration contains now equations (\ref{1})- (\ref{4})
and equations (\ref{27}), (\ref{28}), where  the fluxes $\pi_{ik},
{\bf q}$ and the dissipative function $R$ are given by equations
(\ref{22}), (\ref{26}) and (\ref{21}). As it seen from (\ref{28})
the equation for diffusion flux is the equation of a relaxation
type. At $\tau = 0$, i.e. for slow processes, it determines
directly the diffusion flux in terms of gradients of the basic
hydrodynamic variables.

\section{Mass transfer in field of fast-moving deformation disturbance}

\label{sec:3} In this section we consider the diffusion of
interstitial impurities in the field of the fast-moving
deformation disturbance of the type of a chock wave. We assume
that the stepped disturbance (kink), for which the displacement
field (along the $x$-axis) has the form \cite{K72,PK97}
\begin{equation}
u(x,t) = u_0 \Bigl(1 - \tanh\frac{x - Vt}{\delta}\,\Bigr)\,
,\label{29}
\end{equation}
where $u_0$ is the amplitude of the disturbance and $\delta$ is
the width of the transition region, propagates along the
$x$-direction with the constant velocity $V$ compared with the
velocity of the propagation of concentration disturbances in the
medium.\\
At the given motion of the lattice (\ref{29}) the diffusion
process is largely governed by  the set of two equations: the
local conservation law of mass (\ref{4}) and the relaxation
equation for the diffusion flux (\ref{28}). Let us now transform
the right side of equation (\ref{28}) to the alternative form.
Combining equations (\ref{7}) and (\ref{17}) and taking into
account equation (\ref{20}), one obtains
\begin{equation}
\rho d\mu = - SdT + dp +\nu d\, c + \tsi_{ij}d\tilde{u}_{ij} +
\frac{W}{2}dJ^2 \, , \label{30}
\end{equation}
where $W =\frac{1}{3}W_{ii}$. It follows from  equation (\ref{30})
that one can consider the chemical potential  $\nu$ to be a
function of the independent variables $T, p, c, \tilde{u}_{ij},
J^2$. Omitting essentially nonlinear terms,  note that in the case
of the cubic symmetry a scalar function can depend on the tensor
$\tilde{u}_{ij}$ by convolution $\tilde{u}_{ii}= 0$ only.
Consequently one can consider $\nu$ to be a function of $T, p, c$
only.\\
The available experimental data indicate that when a shock wave
propagates in a a solid the effect of thermodiffusion is neglected
in comparison with borodiffusion \cite{BV99,ZR66}. Here we also
assume that the main factor determining mass transfer in our
system is the effect of barodiffusion. Then omitting the gradient
of a temperature one rewrites equation (\ref{28}) in the more
simple form as
\begin{equation}
\tau\frac{d{\bf J}}{d\,t} + {\bf J} = -\, \gamma T
\frac{\partial}{\partial c}\Bigl(\frac{\nu}{\rho T}\Bigr)\nabla c
- \gamma \frac{\partial}{\partial p}\Bigl(\frac{\nu}{\rho
T}\Bigr)\nabla p - \lambda\nabla\cdot{\bf \tsi}\, , \label{31}
\end{equation}
It follows from equation (\ref{16}) that $\sigma_{ik} = -
p\delta_{ik} + \tsi_{ik}$, with $\sigma_{ii} = - 3p$ is the
elastic stress tensor. In the one-dimensional case one neglects
the influence of shear stress  on a diffusion mobility
\footnote[1]{In the one-dimensional case ${\bf \tsi}$ is
proportional to $\partial u/\partial x$. Therefore  allowance for
the term $\lambda\nabla\cdot{\bf \tsi}$ in the right-hand side of
Eq.(\ref{31}) leads simply to redefinition of the constant $G$ in
Eq. (\ref{33}).}. In addition, we assume that Hook's law $p = -
kU$, relating the pressure $p$ with the deformation $U =
\partial u/\partial x$ of a material element, is valid. Redenoting
the constants in equation (\ref{31}) one writes down in the linear
approximation the coupled set  of the one-dimensional versions  of
equations (\ref{4}) and (\ref{31}) as
\begin{eqnarray}
&&\rho\frac{\partial c}{\partial t} = - \frac{\partial J}{\partial
x}\, , \label{32}\\
&&\tau\frac{\partial J}{\partial t} + J = - \rho D \frac{\partial
c }{\partial x} + \rho G\frac{\partial U}{\partial x}\, ,
\label{33}
\end{eqnarray}
where  $G > 0$ has the meaning of the coefficient of the
barodiffusion and $D = \gamma\rho^{-1}T{\partial (\nu/\rho
T)/{\partial c}}$ is the diffusion constant. In addition, in
obtaining of equation (\ref{32}) and (\ref{33}) we have neglected
the term ${\bf v}\cdot\nabla$, appearing in the material
derivative that is small for fast processes  in comparison with
the time derivative
$\partial /\partial t$.\\
After some algebra, equations (\ref{32}) and (\ref{33}) are
brought to the form
\begin{eqnarray}
&& \frac{\partial c}{\partial t} +
\frac{D}{V^2_D}\frac{\partial^2c}{\partial t^2} - D
\frac{\partial^2 c}{\partial x^2} = - \, G\frac{\partial^2
U}{\partial x^2}\, ,
\label{34}\\
&& \frac{\partial J}{\partial t} +
\frac{D}{V^2_D}\frac{\partial^2J}{\partial t^2} - D
\frac{\partial^2 J}{\partial x^2} = \rho G\frac{\partial^2
U}{\partial x
\partial t}\, . \label{35}
\end{eqnarray}
At $U = 0$ equation (\ref{34}) (as well as equation(\ref{35})) is
known as the telegrapher equation that, unlike to the classical
diffusion equation, gives rise to the finite speed $V^2_D =
D/\tau$ of the propagation  of concentration perturbations
\cite{JL96,S97}.\\
It is convenient to consider equations (\ref{34}) and (\ref{35})
in the reference frame moving with velocity  $V$ together with the
front of the disturbance (\ref{29}). Passing to the dimensionless
coordinate $\xi = (x - Vt)/\tau V_D$, one has
\begin{eqnarray}
&&(\varphi^2 - 1)\frac{d^{2} \bar{c}}{d\xi^2} -
\varphi\frac{d\bar{c}}{d\xi} = - \frac{d^{2} \bar{U}}{d\xi^2}\, ,
\label{36}\\
&&(\varphi^2 - 1)\frac{d^{2} \bar{J}}{d\xi^2} -
\varphi\frac{d\bar{J}}{d\xi} = -\, \varphi\frac{d^{2}
\bar{U}}{d\xi^2}\, , \label{37}
\end{eqnarray}
where $\varphi = V/V_D$, $\bar{U} = GU/c_0D$, $\bar{c} = (c -
c_0)/c_0$, $\bar{J} = J/c_0\rho V_D$ and $c_0$ is the equilibrium
interstitial concentration.\\
Supposing that far from the front of the wave  ($\xi = 0$) the
medium is not disturbed, we take the boundary conditions in the
form
\begin{equation}
\xi \rightarrow\pm\infty,\qquad  \bar{c}\rightarrow 0,\qquad
\bar{J}\rightarrow 0\, . \label{38}
\end{equation}
The solution of the boundary value problem (\ref{36}) - (\ref{38})
can be presented as follows
\begin{equation}
\bar{c}(\xi) = \left\{
\begin {array}{ll}
b\,\bar{U}(\xi) - ab\int\limits_{- \infty}^\xi e^{- a(\xi - \xi ')}\bar{U}(\xi ')d\xi '\, , &\varphi  < 1\, ,  \nonumber \\
\displaystyle{\frac{d\bar{U}}{d\xi}}\, ,&\varphi = 1\, ,  \\
b\,\bar{U}(\xi) + ab\int\limits^{\infty}_\xi e^{- a(\xi - \xi
')}\bar{U}(\xi ')d\xi '\, , &\varphi  > 1\, ,
\end {array}\right.
\label{39}
\end{equation}
Here $a = \varphi b$, $b = 1/(1 - \varphi^2)$ and the solution for
the dimensionless flux $\bar{J}$ follows from (\ref{39}) as a
result of the replacement of $b$ by $a$. Equation (\ref{39}) is
the solution of boundary value problem for arbitrary deformation
disturbance $U$, vanishing at
$\xi\rightarrow\pm\infty$.\\
The transition to the local-equilibrium theory occurs if in
equation (\ref{33}) one puts formally $\tau = 0$. Using the same
dimensionless variables as in equations (\ref{36}) and (\ref{37})
one has in this case the following equations
\begin{eqnarray}
&&\frac{d^{2} \bar{c}_{leq}}{d\xi^2} -
\varphi\frac{d\bar{c}_{leq}}{d\xi} = - \frac{d^{2}
\bar{U}}{d\xi^2}\, ,
\label{41}\\
&&\frac{d^{2} \bar{J}_{leq}}{d\xi^2} -
\varphi\frac{d\bar{J}_{leq}}{d\xi} = -\, \varphi\frac{d^{2}
\bar{U}}{d\xi^2}\, . \label{42}
\end{eqnarray}
The solution of this equations has  the standard form  and we
shall not give it here.\\
For the disturbance (\ref{29}) $ \bar{U}(\xi) = - U_0/(cosh^2\xi
/\delta_1)$,  where $U_0$ is the known constant and $\delta_1 =
\delta / \tau V_D$, one rewrites equation (\ref{39}) in the form
\begin{equation}
\bar{c}(\bar{\xi})/U_0 = \left\{
\begin {array}{ll}
\displaystyle - \frac{b}{\cosh^2\bar{\xi}} +
(a\delta_1)b\int\limits_{- \infty}^{\bar{\xi}}\frac{ e^{- a\delta_1(\bar{\xi} - \xi )}}{\cosh^2\xi}\, d\xi \, , &\varphi  < 1\, ,  \nonumber \\
\displaystyle{\frac{2}{\delta_1}\frac{\sinh\bar{\xi}}{\cosh^2\bar{\xi}}} \, ,&\varphi = 1\, ,  \\
\displaystyle - \frac{b}{\cosh^2\bar{\xi}} -
(a\delta_1)b\int\limits_{\bar{\xi}}^{\infty}\frac{e^{-
a\delta_1(\bar{\xi} - \xi)}}{\cosh^2\xi}\, d\xi \, , &\varphi
> 1\, ,
\end {array}\right.
\label{46}
\end{equation}
here $\bar{\xi} = \xi/\delta_1$. Equation (\ref{46}) is the
solution of the given problem at arbitrary values of parameters.
Here we are interested in values of the parameter $\varphi =
V/V_D$ close to one for which $|a| >> 1,\, |b|>> 1$. At the same
time the parameter $\delta_1$, characterizing the  width of the
transition region of the disturbance, can take, generally
speaking, any values. In what follows we confine
ourselves to two limiting cases. \\
 1.\textit{The wide transition region}, $\delta_1 =
\delta /\tau V_D \sim 1$. Figures \ref{fig1}-\ref{fig3} present
the curves for the impurity concentration $\bar{c}/U_0$ (as well
as for the diffusion flux $\bar{J}/\varphi U_0$) at different
values of the parameter $\varphi$ in the case of the propagating
local compression ($U_0 > 0$). This figures also show the
difference between the local equilibrium and the local
nonequilibrium (presence of the relaxation process in the medium)
theories. In all cases the deformation disturbance produces the
enhanced impurity concentration immediately  before the front of
the deformation wave (the region of $\bar{\xi} > 0$), lowering it
behind the front. In the region of the of the enhanced
concentration the diffusion flux points in the direction of the
disturbance propagation, while in the region of lowered
concentration - in the opposite direction.\\
As figures \ref{fig2} and \ref{fig3} show, at $\varphi\approx 1$
the concentration perturbations are significantly greatly in the
local nonequilibrium medium. In this case the diffusion flux is
nearly twice as large as that for the local equilibrium system.\\
2.\textit{The narrow transition region}, $\delta_1 = \delta /\tau V_D
<< 1$. In what follows we consider the special case of $\delta_1 =
1/|a|,\quad |a| >> 1$. Figures \ref{fig4}-\ref{fig5} demonstrate a
pronounced difference between two approaches. The ejection of the major
number of interstitials into the region before the front of the wave is
observed in the absence of a local equilibrium in the medium. In the
latter case the impurity concentration in the indicated region is
nearly  by an order greater than the local equilibrium concentration
that does not change practically with the rising of $\varphi$. Figures
\ref{fig4}-\ref{fig5} also show the considerable deviation for
diffusion fluxes demonstrating the acceleration of the mass transfer
under local nonequilibrium conditions. Note that at $\varphi > 1$ in
the region of the front the diffusion fluxes referring to the different
nonequilibrium states of the medium point in complete opposite
directions. The difference between the local nonequilibrium curves at
different
magnitudes $\varphi$ are shown in figure \ref{fig6}.\\
The reason of the enhanced impurity concentration before the front in
the local-nonequilibrium system is that when the velocity of the
deformation wave is near to $V_D$, the concentration disturbances have
no time to propagate on large distances from the front. This has the
effect of accumulation of interstitials in the indicated region. Here
is a complete analogy with the task about the motion of a thermal
source in a relaxating medium \cite{S91}.\\
It should be noted that in the case of the local tension ($U_0 < 0$)
the concentration peaks will form behind the front of the wave
($\bar{\xi} < 0$) by the moving particles from the region $\bar{\xi} >
0$.

\section{Summary and conclusion}
\label{sec:4}

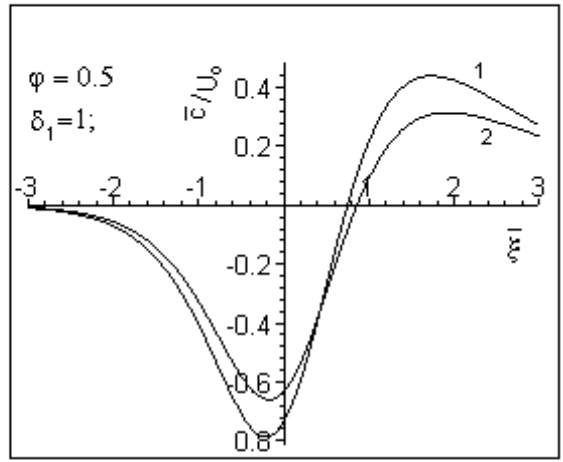
\begin{figure}
\caption{\label{fig1}Normalized interstitial concentration
$\bar{c} /U_0$
 (normalized dimensionless diffusion flux $\bar{J}/\varphi U_0$) vs.
 dimensionless coordinate $\bar{\xi}$ at $\varphi = 0.5$, $\delta_1 =
 1$ and $U_0 > 0$. Local nunequilibrium system - curve 1; local equilibrium system - curve 2.}
 \end{figure}

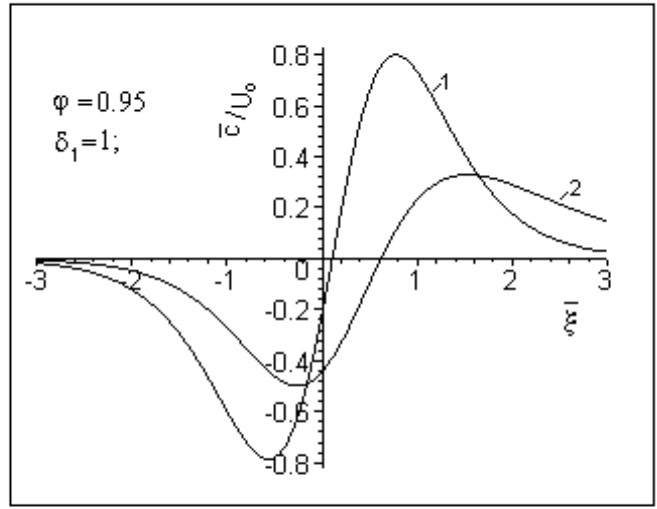
\begin{figure}
\caption{\label{fig2}Normalized interstitial concentration
$\bar{c} /U_0$
 (normalized dimensionless diffusion flux $\bar{J}/\varphi U_0$) vs.
 dimensionless coordinate $\bar{\xi}$ at $\varphi = 0.95$, $\delta_1 =
 1$ and $U_0 > 0$. Local nunequilibrium system - curve 1; local equilibrium system - curve 2.}
 \end{figure}

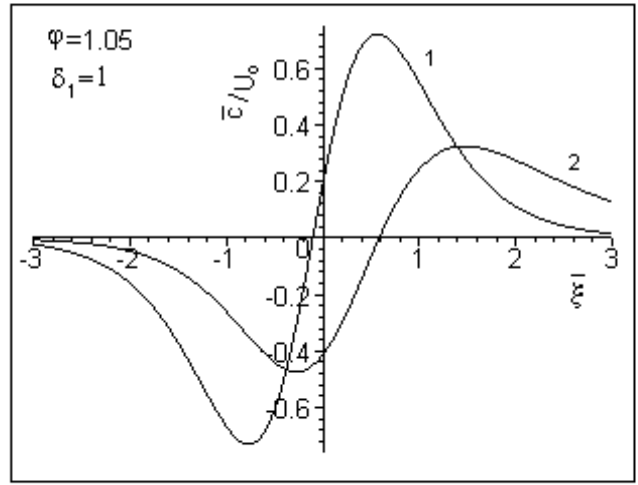
\begin{figure}
 \caption{\label{fig3}Normalized interstitial concentration $\bar{c} /U_0$
 (normalized dimensionless diffusion flux $\bar{J}/\varphi U_0$) vs.
 dimensionless coordinate $\bar{\xi}$ at $\varphi = 1.05$, $\delta_1 =
 1$ and $U_0 > 0$. Local nunequilibrium system - curve 1;
 local equilibrium system - curve 2.}
 \end{figure}

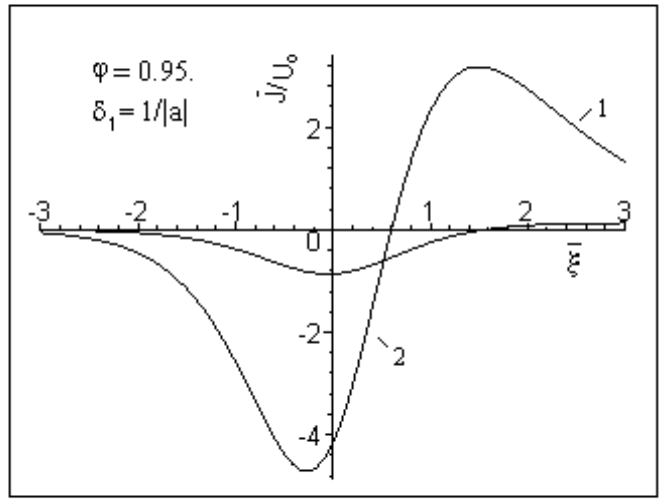
\begin{figure}
 \caption{\label{fig4}Normalized dimensionless diffusion flux ${\bar{J}/U_0}$
 (normalized interstitial concentration $\varphi \bar{c}/U_0$)
 vs. dimensionless coordinate $\bar{\xi}$ at $\varphi = 0.95$,
 $\delta_1 = 1/|a|$ and $U_0 > 0$. Local nonequilibrium
 system - curve 1; local equilibrium - curve 2.}
 \end{figure}

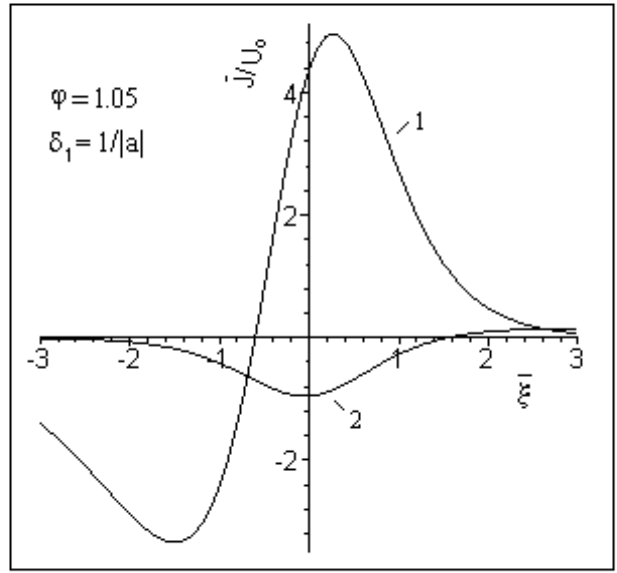
\begin{figure}
 \caption{\label{fig5}Normalized dimensionless diffusion flux ${\bar{J}/U_0}$
 (normalized interstitial concentration $\varphi \bar{c}/U_0$)
 vs. dimensionless coordinate $\bar{\xi}$ at $\varphi = 1.05$,
 $\delta_1 = 1/|a|$ and $U_0 > 0$. Local nonequilibrium
 system - curve 1; local equilibrium - curve 2.}
 \end{figure}

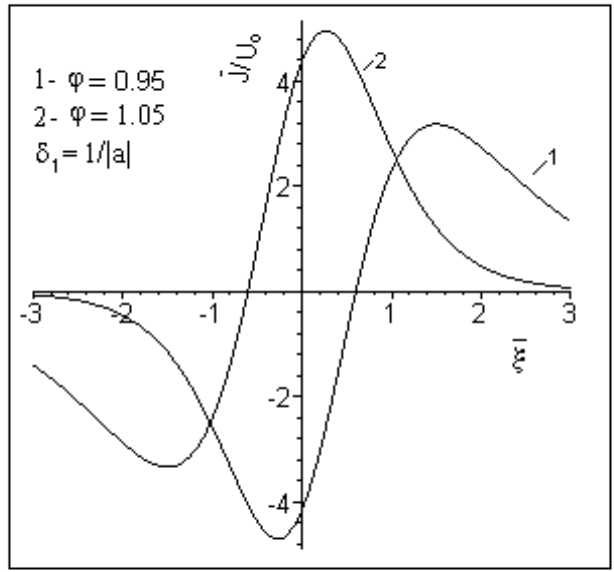
\begin{figure}
 \caption{\label{fig6}Normalized dimensionless diffusion flux $\bar{J}/U_0$
 (normalized interstitial concentration $\varphi \bar{c}/U_0$) vs.
 dimensionless coordinate $\bar{\xi}$ for the case of local
 nonequilibrium system at $\delta_1 = 1/|a|$, $U_0 > 0$ and $\varphi = 0.95$ (curve 1),
 $\varphi = 1.05$ (curve 2).}
 \end{figure}

In this work we have presented the investigation of mass transfer
in the field of the fast-moving deformation disturbance of the
type of a shock wave, propagating in the infinite lattice with
interstitial impurities. The research is motivated by experiments
on the exposure of solid to dynamical external loads
\cite{K94,BV99,BK99,P90,BG83,BC86,NT86,MP84,NV85}\\
For the description of fast hydrodynamical processes the extending
of the set of the conventional variables, characterizing the
nonequilibrium state  of a solid is needed. In the given model the
only extra variable is the diffusion flux, the relaxation time of
which is assumed to be comparable to the timescale of the process
under consideration. For the selected variables the  set
of the coupled hydrodynamical equations has been derived. \\
Having assumed that the main factor impacting on mass transfer in
our case is the effect of barodiffusion \cite{BV99,ZR66}, we
reduce the diffusion task to the investigation of the set of two
linear, one-dimensional equations. These are the equation for the
interstitial concentration (\ref{36}) and the relaxation equation
for the diffusion flux (\ref{37}). The solution of this set is
defined by two dimensionless parameters $\varphi = V/V_D$ and
$\delta_1 = \delta/\tau V_D$. At the different magnitudes $\varphi
\sim 1$ we have considered two special cases $\delta_1 \simeq 1$
and $\delta_1 << 1$.\\
Figures \ref{fig1}-\ref{fig5} show that the occurrence of the
relaxation process in the medium leads to the significant
deviation of the concentration profiles and the diffusion fluxes
from the local equilibrium diffusion data.  The ejection of
interstitials into the region before the front enables to speak
about the effect of the entrainment of impurity particles by the
wave.  This effect takes place in the local equilibrium system
also (see \cite{S00} as well), however at $\delta_1 << 1$ to a
largest measure it is displayed in the local
nonequilibrium conditions (figures \ref{fig4} and \ref{fig5}).\\
In the case of the local tension ($U_0 < 0$; the mirror reflection
of the plots \ref{fig1}-\ref{fig6} relative to axis $\bar{\xi}$)
the concentration peaks will form beyond the front of the wave by
a "suction" of particles from the range before the front into the
extended range.\\
The case of $\delta_1 \sim 1$ ($\delta \sim \tau V_D$) corresponds
to the smooth enough behavior of the deformation disturbance.
Here, in the region of the front, the local nonequilibrium
diffusion flux is nearly twice as large as that for the local
equilibrium system. The case of $\delta_1 << 1$ ($\delta << \tau
V_D$) corresponds to the sharp deformation variations. From
figures \ref{fig4} and \ref{fig5} it is seen that in this case one
can speak about the significant acceleration of the mass transfer
in the local nonequilibrium system. The all effects noted above
are by an order
less in the local equilibrium system.\\
Thus our investigation has shown that at the taken magnitudes of
the parameters of the model an enhanced mass transfer takes place
in local nonequilibrium system in comparison with the same process
considered under the local equilibrium conditions.\\
Note in conclusion that the results presented here have been
obtained within the scope of the linear theory.  The set of the
hydrodynamical equations derived in section \ \ref{sec:2} can form
the basis for the investigation of nonlinear effects.

\newpage

\end{document}